\documentclass[10pt]{emulateapj}
\usepackage{epsfig}
\usepackage{amsmath}
\usepackage{natbib}
\makeatletter
\newcommand*{\rom}[1]{\expandafter\@slowromancap\romannumeral #1@}
\makeatother

\def\ion#1#2{\ifmmode \mbox{{\rm #1}}\,\mbox{{\sc #2}} 
        \else {\rm #1}$\,${\sc #2}
        \fi}
\def\CI{\ion{C}{i}}

\def\NII{\ion{N}{ii}}

\begin{document}

\title{Evidence for CO shock excitation in NGC 6240 from Herschel SPIRE spectroscopy}

\author{R. Meijerink$^{1,2}$} 
\author{L.E. Kristensen$^{2}$} 
\author{A. Wei$\ss^3$} 
\author{P.P. van der Werf$^2$}
\author{F. Walter$^4$}
\author{M. Spaans$^1$}
\author{A.F. Loenen$^2$}
\author{J. Fischer$^5$}
\author{F.P. Israel$^2$}
\author{K. Isaak$^6$}
\author{P.P. Papadopoulos$^3$}
\author{S. Aalto$^7$}
\author{L. Armus$^8$}
\author{V. Charmandaris$^9$}
\author{K.M. Dasyra$^{10}$}
\author{T. Diaz-Santos$^8$} 
\author{A. Evans$^{11,12}$}
\author{Y. Gao$^{13}$}
\author{E. Gonz\'alez-Alfonso$^{14}$}
\author{R. G\"usten$^3$}
\author{C. Henkel$^{3,15}$}
\author{C. Kramer$^{16}$}
\author{S. Lord$^{17}$}
\author{J. Mart\'in-Pintado$^{18}$}
\author{D. Naylor$^{19}$}
\author{D.B. Sanders$^{20}$}
\author{H. Smith$^{21}$}
\author{L. Spinoglio$^{22}$}
\author{G. Stacey$^{23}$}
\author{S. Veilleux$^{24}$}
\author{M.C. Wiedner$^{25}$}

\address{$^1$ Kapteyn Astronomical Institute, PO Box 800, 9700 AV
  Groningen, The Netherlands}
\address{$^2$ Leiden Observatory, Leiden
  University, P.O. Box 9513, NL-2300 RA Leiden, Netherlands}
\address{$^3$ Max-Planck-Institut f\"ur Radioastronomie, Auf dem H\"ugel
  16, Bonn, D-53121, Germany}
\address{$^4$ Max-Planck-Institut f\"ur Astronomie, K\"onigstuhl 17,
  Heidelberg, D-69117, Germany}
\address{$^5$ Naval Research Laboratory, Remote Sensing Division,
  Washington, DC 20375, USA}
\address{$^6$ ESA Astrophysics Missions Division, ESTEC, PO Box 299,
  2200 AG Noordwijk, The Netherlands}
\address{$^7$ Department of Radio and Space Science, Onsala Observatory,
  Chalmers University of Technology, 43992 Onsala, Sweden}
\address{$^8$ Spitzer Science Center, California Institute of Technology,
MS 220-6, Pasadena, CA 91125, USA}
\address{$^9$ University of Crete, Department of Physics, 71003
  Heraklion, Greece}
\address{$^{10}$Observatoire de Paris, LERMA (CNRS:UMR8112), 61 Av. de
  l'Observatoire, F-75014, Paris, France} 
\address{$^{11}$ Department of Astronomy, University of Virginia, 530 McCormick
Road, Charlottesville, VA 22904, USA}
\address{$^{12}$ National Radio Astronomy Observatory, 520 Edgemont Road,
Charlottesville, VA 22903, USA}
\address{$^{13}$Purple Mountain Observatory, Chinese Academy of Sciences, 2
  West Beijing Road, Nanjing 210008, PR China}
\address{$^{14}$ Universidad de Alcal\'a Henares, Departamente de F\'isica, Campus
Universitario, 28871 Alcal\'a de Henares, Madrid, Spain}
\address{$^{15}$Astron. Dept.,
  King Abdulaziz University, P.O. Box 80203, Jeddah, Saudi Arabia}
\address{$^{16}$ Instituto Radioastronomie Millimetrica (IRAM), Av. Divina
Pastora 7, Nucleo Central, 18012 Granada, Spain}
\address{$^{17}$ NASA Herschel Science Center, California Institute of
  Technology, M.S. 100–22, Pasadena, CA 91125, USA}
\address{$^{18}$Departamento de Astrofisica Molecular e Infrarroja-Instituto de
Estructura de la Materia-CSIC, Calle Serrano 121, 28006 Madrid,
Spain}
\address{$^{19}$Department of Physics, University of Lethbridge, 4401
  University Drive, Lethbridge, Alberta, T1J 1B1, Canada}
\address{$^{20}$ University of Hawaii, Institute for Astronomy, 2680 Woodlawn
Drive, Honolulu, HI 96822, USA}
\address{$^{21}$ Harvard-Smithsonian Center for Astrophysics, 60 Garden Street,
Cambridge, MA 02138, USA}
\address{$^{22}$ Istituto di Astrofisica e Planetologia Spaziali, INAF, Via Fosso del
Cavaliere 100, I-00133, Roma, Italy}
\address{$^{23}$ Department of Astronomy, Cornell University, Ithaca, NY 14853, USA}
\address{$^{24}$ Department of Astronomy, University of Maryland, College Park,
MD 20742, USA}
\address{$^{25}$ Observatoire de Paris, LERMA, CNRS, 61 Av. de l'Observatoire, 75014
Paris, France}

\begin{abstract} 
  We present Herschel SPIRE FTS spectroscopy of the nearby luminous
  infrared galaxy NGC 6240. In total 20 lines are detected, including
  CO $J=4-3$ through $J=13-12$, 6 H$_2$O rotational lines, and [$\CI$]
  and [$\NII$] fine-structure lines. The CO to continuum luminosity
  ratio is 10 times higher in NGC 6240 than Mrk 231. Although the CO
  ladders of NGC 6240 and Mrk 231 are very similar, UV and/or X-ray
  irradiation are unlikely to be responsible for the excitation of the
  gas in NGC 6240. We applied both C and J shock models to the H$_2$
  $v=1-0\ S(1)$ and $v=2-1\ S(1)$ lines and the CO rotational
  ladder. The CO ladder is best reproduced by a model with shock
  velocity $v_s=10$~km\,s$^{-1}$ and a pre-shock density $n_{\rm
    H}=5\times 10^4$~cm$^{-3}$.  We find that the solution best
  fitting the H$_2$ lines is degenerate: The shock velocities and
  number densities range between $v_s = 17 - 47$~km\,s$^{-1}$ and
  $n_{\rm H}=10^7 - 5\times 10^{4}$~cm$^{-3}$, respectively. The H$_2$
  lines thus need a much more powerful shock than the CO lines. We
  deduce that most of the gas is currently moderately stirred up by
  slow (10 km\,s$^{-1}$) shocks while only a small fraction ($\lesssim
  1$ percent) of the ISM is exposed to the high velocity shocks. This
  implies that the gas is rapidly loosing its highly turbulent
  motions. We argue that a high CO line-to-continuum ratio is a key
  diagnostic for the presence of shocks.
\end{abstract}



\section{Introduction}

We present {\it Herschel}\footnote{Herschel is an ESA space
  observatory with science instruments provided by European-led
  Principal Investigator consortia and with important participation
  from NASA} SPIRE FTS \citep{Griffin2010} observations of the nearby
luminous infrared galaxy NGC 6240 (IRAS 16504+0228, UGC 10592). With a
redshift of $z=0.024576$ and a 7-year WMAP flat cosmology ($H_0 =
70.50$~km\,s$^{-1}$\,Mpc$^{-1}$, $\Omega_{\rm matter} = 0.27$,
$\Omega_{\rm vacuum} = 0.73$) NGC 6240 is at a luminosity distance of
$D_{\rm L} = 107$~Mpc, with $1\arcsec$ corresponding to 492 pc. The
derived $8-1000$~$\mu$m luminosity of the merger galaxy NGC 6240 is
$L_{\rm IR} = 7.5\times 10^{11}$~L$_{\odot}$. Different power sources
are suggested for this infrared luminosity, which complicate the
interpretation of observations. The source is a strong X-ray
emitter. Based on modeling of Chandra between $0.5-8$~keV,
\citet{Komossa2003} derive extinction corrected luminosities of
$L_{{\rm X},N}=1.9\times10^{42}$~erg~s$^{-1}$ and $L_{{\rm
    X},S}=0.7\times10^{42}$~erg~s$^{-1}$ for the northern and southern
AGN cores in the energy range $0.1-10$~keV. However, using BeppoSAX
observations, \citet{Vignati1999} find that the intrinsic hard X-ray
luminosity shows up above $9-10$~keV, and their models of the higher
energy yields $L(2-10\ {\rm keV})\approx 3.6 \times 10^{44}$~erg
s$^{-1}$. Hydrogen recombination lines \citep{Rieke1985, Depoy1986,
  Elston1990, VdWerf1993} and luminous PAH emission \citep{Armus2006}
in the two nuclei indicate recent star formation. The H$_2$ $v=1 - 0\
S(1)$ 2.12 $\mu$m line, presented by \citet{VdWerf1993}, shows a peak
in the overlap region between these two nuclei, which are 2$\arcsec$
apart. This H$_2$ emission extends over several kpc and shows a
complex morphology. The authors conclude that the bright H$_2$
emission between the nuclei is generated in shocks resulting from the
collision of the interstellar media (ISM) of the merging
galaxies. More recently, \citet{Engel2010} observed the same line at
high spatial (0.5\arcsec) and spectral ($\sim 90$~km~s$^{-1}$)
resolution. The line profiles also indicate multiple components, and
the dispersion map suggests that the gas is highly disturbed and
turbulent. Interferometric observations of several CO, HCN, and
HCO$^+$ lines show a high-density peak between the two nuclei
\citep{Tacconi1999, Nakanishi2005, Iono2007}. \citet{Tacconi1999}
conclude that NGC 6240 is in an earlier merging stage than Arp 220,
the prototypical ultra-luminous infrared galaxy, and state that the
gas is in the process of settling between the two nuclei, and
dissipating angular momentum rapidly. \citet{Engel2010} also presented
a $^{12}$CO $J=2-1$ interferometric map, and conclude that the H$_2$
and CO emission are coextensive, but do not coincide with the stellar
emission distribution.

This paper is ordered as follows. In Section 2, the observations, data
reduction and line luminosities are discussed. In Section 3, we
discuss the different excitation components (PDR, XDR, shocks) in
combination with the geometry of NGC 6240. Consequently, we compose a
CO ladder using both the available SPIRE FTS and ground based CO data
from \citet{Papadopoulos2011} and compare this system to the ULIRG Mrk
231 \citep[previously studied by us with {\it
  Herschel},][]{VdWerf2010}. Then we analyze this with shock models
from \citet{Kristensen2007}. In Section 4, we conclude with a
discussion and a summary of the results.

\section{Observations, data reduction, and results}

NGC 6240 was observed (Observations ID 1342214831) in staring
mode with the SPIRE FTS on February 27, 2011, as part of the {\it
  Herschel} Open Time Key Program HerCULES (P.I. Van der Werf). The
high spectral resolution mode was used, yielding a resolution of
$1.2$~GHz over both observing bands: the low frequency band covering
$\nu = 447 - 989$~GHz ($\lambda = 671 - 303$~$\mu$m) and the high
frequency band covering $\nu = 958 - 1545$~GHz ($\lambda = 313 -
194$~$\mu$m). In total 97 repetitions (194 FTS scans) were carried
out, yielding an on source integration time of 12920~s (3.6 hrs) for
each band. A reference measurement was used to subtract the combined
emission from the sky, telescope and instrument. The data were
processed and calibrated using HIPE version 6.0. The extent of the CO
$J=3-2$ emission is $4\arcsec$ \citep{Wilson2008}, while the SPIRE
beam varies from $17\arcsec$ to 42$\arcsec$ over our
spectrum. Therefore, a point source calibration procedure was adopted,
and no corrections for wavelength dependent beam coupling factors were
necessary.

\begin{figure*}
  \centering
  \includegraphics[width=15cm,clip]{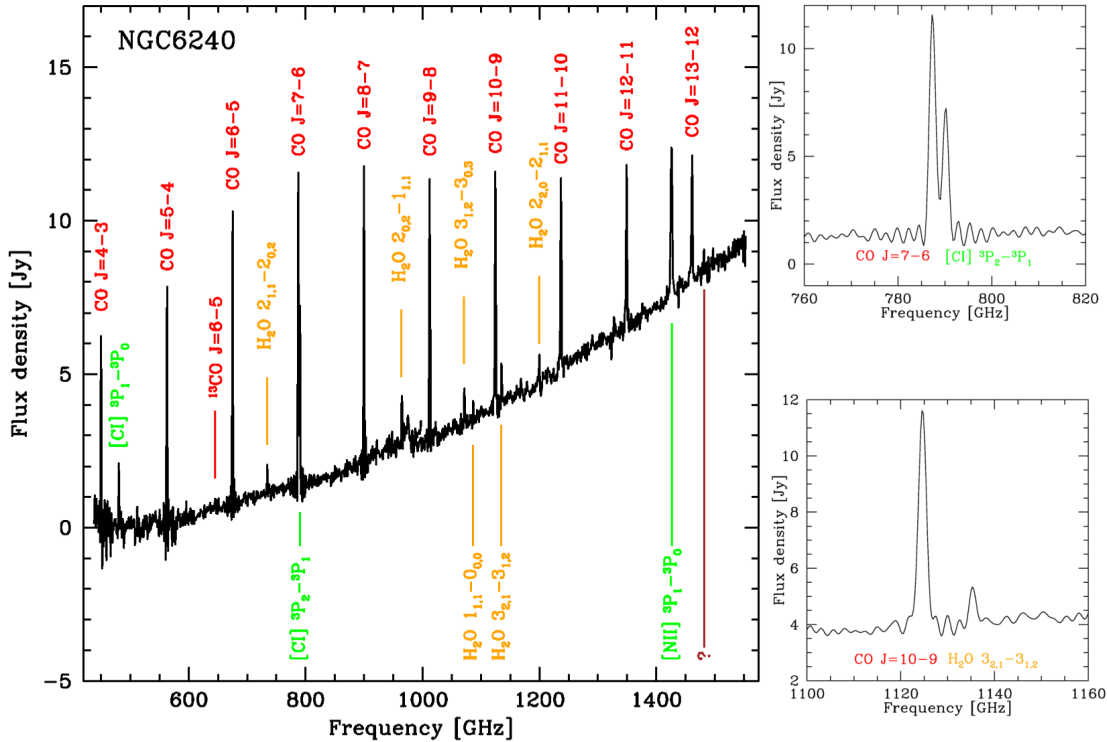}
  \caption{Full SPIRE FTS spectrum of NGC~6240 and zoom-in on line
    blends, with on the x-axis the frequency and on the y-axis the
    flux density in Jy. The detected lines are marked: CO (red),
    [$\CI$] and [$\NII$] (green), H$_2$O (orange), and unidentified
    (brown). In the overlap region between the two frequency bands
    ($958 - 989$~GHz), the noisy parts were clipped and plotted on top
    of each other.}
  \label{SPIRE_NGC6240}
\end{figure*}

The full SPIRE FTS spectrum of NGC 6240 is shown in
Fig. \ref{SPIRE_NGC6240}. In the overlap region between the two
frequency bands ($958 - 989$~GHz), the noisy parts of the two
spectrometer bands were clipped and plotted on top of each other. A
total of 20 lines were detected of which one line at the observed
frequency $\nu_{\rm rest}=1481.6$~GHz remains unidentified. We also
see hints of absorption and emission between, e.g., 1300 and 1350
GHz. The significance of these features are still under investigation,
and requires a more advanced flux extraction than currently
adopted. The identified lines include the CO $J=4-3$ to $J=13-12$, 6
H$_2$O lines, [$\CI$] 370 and 609~$\mu$m, and [$\NII$] 205~$\mu$m. A
well-constrained upper limit was obtained for the $^{13}$CO $J=6-5$
line. The line fluxes are listed in Table \ref{Fluxes}. In all cases a
simple integration over the line was done. This procedure was repeated
with approximately 100 different choices of baseline and integration
intervals. Specifically, the integration is done from $\pm 5$ to
$\pm9$~GHz centered on the line with steps of 0.4~GHz (10 integration
borders). For each of these integration borders, the baseline window
is set from $\pm 2$ to $\pm3.5$~GHz with steps of 0.15~Ghz (10
iterations for each of the integration borders). The average is taken
as the line flux and the standard deviation of the different
realizations is taken as the error, and excludes the relative
calibration error of 7 percent. This is also done for fluxes of lines
that are blended. For these, for each iteration, a double Gaussian was
fitted and the peaks were used to set the fractional flux for each of
the lines. As additional modeling constraints, the table lists fluxes
for three lower CO transitions \citep{Papadopoulos2011}, and
luminosities of four H$_2$ transitions \citep{Tecza2000}. From these
four H$_2$ lines, \citet{Tecza2000} estimated an unattenuated power of
$L_{\rm H_2}\approx 2\times 10^9$~L$_\odot$.

\begin{table}
  \caption[]{Derived fluxes from the SPIRE FTS observations supplemented with ground-based observations}
 
\label{Fluxes}    
\begin{center}
\begin{tabular}{l c c c c}    
\vspace{-5.0mm}    \\
\hline\hline               
\noalign{\smallskip}
Line         & $\nu_{\rm obs}$ & $\lambda_{\rm rest}$  & $S_{\rm line}$     & $L_{\rm line}$ \\  
             & [Ghz]            & [$\mu$m]           & [Jy km s$^{-1}$]  & [L$_\odot$]  \\

\noalign{\smallskip}
\hline               
\noalign{\smallskip}
CO $J=1-0^a$            & & 2600.76    & $322\pm29$   & $4.3\times 10^{5,d}$ \\
CO $J=2-1^a$            & & 1300.40    & $1490\pm250$ & $4.0\times 10^6$ \\
CO $J=3-2^a$            & & 866.963    & $3210\pm640$ & $1.2\times 10^7$ \\
CO $J=4-3$              & 449.99 & 650.252    & $4630\pm370$ & $2.8\times 10^7$ \\
CO $J=5-4$              & 562.41 & 520.231    & $5640\pm150$ & $3.8\times 10^7$ \\
CO $J=6-5$              & 674.83 & 433.556    & $5910\pm82$  & $4.8\times 10^7$ \\
CO $J=7-6$              & 787.25 & 371.650    & $6010\pm60$  & $5.6\times 10^7$ \\ 
CO $J=8-7$              & 899.68 & 325.225    & $5830\pm89$  & $6.3\times 10^7$ \\ 
CO $J=9-8$              & 1012.1 & 289.120    & $4770\pm82$  & $5.7\times 10^7$ \\ 
CO $J=10-9$             & 1124.5 & 260.240    & $4160\pm67$  & $5.6\times 10^7$ \\ 
CO $J=11-10$            & 1236.9 & 236.613    & $3160\pm75$  & $4.7\times 10^7$ \\ 
CO $J=12-11$            & 1349.1 & 216.927    & $2590\pm60$  & $4.2\times 10^7$ \\ 
CO $J=13-12$            & 1461.2 & 200.272    & $2080\pm60$  & $3.6\times 10^7$ \\
$^{13}$CO $J=6-5^b$       & 645.21 & 453.498     & $<112\pm67$   & $8.6\times 10^5$ \\
$[\CI]$ $^3 P_1 - {^3}P_0$ & 480.27 & 609.135  & $1750\pm220$ & $1.0\times 10^7$ \\
$[\CI]$ $^3 P_2 - {^3}P_1$ & 789.95 & 370.415  & $3330\pm52$  & $3.1\times 10^7$ \\
$[\NII]$ $^3 P_1 - {^3}P_0$ & 1426.1 & 205.300 & $3220\pm77$  & $5.5\times 10^7$ \\
H$_2$O $2_{11}-2_{02}$	& 734.22 & 398.643    & $522\pm52$   & $4.8\times 10^6$ \\
H$_2$O $2_{02}-1_{11}$	& 964.61 & 303.456    & $798\pm75$   & $9.2\times 10^6$ \\
H$_2$O $3_{12}-3_{03}$	& 1071.5 & 273.193    & $548\pm60$   & $7.0\times 10^6$ \\
H$_2$O $1_{11}-0_{00}$	& 1086.7 & 269.272    & $234\pm52$   & $3.0\times 10^6$ \\
H$_2$O $3_{21}-3_{12}$	& 1135.3 & 257.795    & $606\pm60$   & $8.2\times 10^6$ \\
H$_2$O $2_{20}-2_{11}$	& 1199.8 & 243.974    & $394\pm60$   & $5.6\times 10^6$ \\
UID line                & 1472.7 & 198.689    & $277\pm60$   & $4.9\times 10^6$ \\
H$_2$ $v=1-0\ S(1)^c$    & & 2.12       &              & $3.5 \times 10^7$ \\ 
H$_2$ $v=1-0\ S(0)^c$   & & 2.22       &              & $1.1 \times 10^7$ \\ 
H$_2$ $v=2-1\ S(1)^c$   & & 2.25       &              & $4.9 \times 10^6$ \\ 
H$_2$ $v=2-1\ S(2)^c$   & & 2.15       &              & $2.2 \times 10^6$ \\ 
\vspace{-1.0mm}    \\
\hline
\vspace{-5.0mm}    \\
\end{tabular}
\end{center}
$^a$\citet{Papadopoulos2011} \\
$^b$upper limit \\
$^c$\citet{Tecza2000}, extinction corrected values, luminosities adjusted to adopted distance \\
$^d$relative uncertainties for $L_{\rm line}$ are the same as for $S_{\rm line}$ 
\end{table}

\section{Analysis}

We will focus on the excitation of the CO ladder, and compare CO
excitation conditions to those of H$_2$. A multicomponent Large
Velocity Gradient (LVG) analysis of the CO ladder, considering also
constraints from the HCN and [CI] lines will be presented in another
paper (Papadopoulos et al., in prep.). Preliminary mass estimates from
this study are consistent with those found by
\citet{Papadopoulos2012}, \citet{Greve2009}, \citet{Tacconi1999}, and
\citet{Engel2010}. Estimates range between $M_{\rm total}=3\times
10^{9}$ and $4\times 10^{10}$~M$_{\odot}$, depending on whether low
and/or high density tracers are used.

\begin{figure}
  \centering
  \includegraphics[width=8cm,clip]{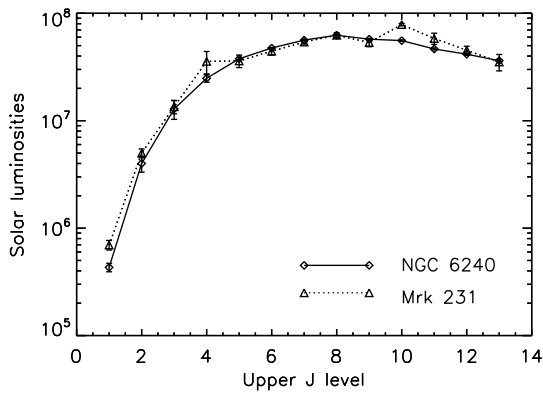}
  \caption{Comparison of the CO ladder of NGC 6240 to those obtained
    for Mrk 231 \citep{VdWerf2010}. The CO ladder of Mrk 231 is
    normalized to the CO $J=8-7$ luminosity of NGC 6240.}
  \label{comparison}
\end{figure}

The total luminosity in the $^{12}$CO lines listed in Table
\ref{Fluxes} is $L_{\rm CO} \approx 5\times 10^{8}$~L$_\odot$. In the
absence of $J_{\rm up} > 13$ measurements this is a lower limit to the
total luminosity in the CO lines. Although the line luminosity of the
CO $J=8-7$ transition is largest, the luminosity in the individual CO
lines is only slowly decreasing for higher rotational quantum numbers,
and a significant contribution is expected from transitions with
$J_{\rm up} > 13$. The shape of the CO ladder of NGC 6240 is similar
to that of Mrk 231 (see Fig. \ref{comparison}) and can be fitted by
two photon-dominated region (PDR) models and an X-ray dominated region
(XDR) model \citep[see][]{VdWerf2010}. The physical and geometrical
properties of NGC 6240 are different from those of Mrk 231. We argue
below why the fit used for Mrk 231 is not appropriate for NGC 6240,
and why shocks must be responsible for the CO excitation in NGC 6240:

{\it Absence of OH$^+$ and H$_2$O$^+$:} The NGC 6240 spectrum does not
show emission line features of the ionic species OH$^+$ and
H$_2$O$^+$, observed in Mrk 231 \citep{VdWerf2010}. Large OH$^+$ and
H$_2$O$^+$ abundances are only sustained in gas clouds with high
ionization fractions ($x_e > 10^{-3}$), which are produced by elevated
cosmic ray or X-ray fluxes \citep[cf.,][]{Meijerink2011}. Their
absence hints that the bulk of the gas is not exposed to high
ionization rates resulting from AGN or starburst/supervae activity.

{\it Line-to-continuum ratio:} The CO luminosity (up to the $J=13-12$
transition) to infrared (measured between $8-1000$~$\mu$m) luminosity
ratio is $L_{\rm CO}/L_{\rm IR}=7\times 10^{-4}$ in NGC 6240. This is
exceptionally high, and approximately an order of magnitude higher
than the ratio found for Mrk 231 \citep{VdWerf2010} {and Arp 220
  \citep{Rangwala2011}}. An exceptionally high line to FIR continuum
ratio is also found for the H$_2$ lines of NGC 6240
\citep{VdWerf1993}. Our PDR and XDR models
\citep{Meijerink2005,Meijerink2007} give a maximum $L_{\rm CO} /
L_{\rm IR}$ ratio of $\sim10^{-4}$ \citep[assuming $I({\rm
  FIR})=2.6\times 10^{-4}G_0$ erg~s$^{-1}$~cm$^{-2}$~sr$^{-1}$,
see][]{Kaufman1999}, where the XDR ratios are highest. Most of the
absorbed photons in a PDR will heat the dust. An AGN (creating an XDR)
generates a UV continuum which contains approximately ten times more
energy than the X-ray field, and also heats the dust efficiently. In
shocks, on the contrary, the gas is compressed and heated to higher
temperatures, while the dust is not affected (except for shock
velocities and densities that are orders of magnitude higher).
Assuming that shocks are not heating the dust and that all the
far-infrared luminosity is reprocessed radiation from the AGN, we
obtain a maximum AGN contribution of $10-15$ percent. So, a shock
dominated ISM can yield a much larger line-to-continuum ratio than
PDRs and XDRs and this is exactly what we see in NGC 6240.

{\it Geometry of NGC 6240:} The bulk of the gas mass does not coincide
with the two AGN nuclei or with star formation. \citet{Engel2010}
relate gas masses, as traced by the CO $J=2-1$ emission, to the
different locations in NGC 6240: $5\%$ and 25$\%$ to the northern and
southern AGN nuclei, and $70\%$ to the CO peak in the overlap region,
while using apertures of 1\arcsec. Our FTS beam ($>19\arcsec$) is
larger than the galaxy, and traces the total CO emission. We
determined a FWHM of $450\pm40$~km~s$^{-1}$ for the CO $J=13-12$
emission with a gaussian$\times$sinc profile fit (larger than the
instrumental resolution of 245~km~s$^{-1}$ at this frequency
range). This is similar to the FWHM of the CO $J=2-1$ line at the
emission peak between the two nuclei, suggesting that the CO $J=13-12$
emission is also located in this region. The projected distance
between these nuclei on the sky is $\sim750$~pc, and the true distance
between them is estimated at 1.4 kpc \citep{Tecza2000}. These authors
assume that the two nuclei are on a circular orbit, of which the
position angle and inclination are the same as that of the CO-disk
between the two nuclei \citep{Tacconi1999}, and that the velocity
difference between the two nuclei is 150 km~s$^{-1}$. The AGN X-ray
luminosities are not enough to power the CO excitation of the gas that
is residing between the two nuclei, since they are geometrically
diluted and absorbed: The luminosities derived by \citet{Komossa2003}
give an X-ray flux at 250 pc from the AGN of $<
1$~erg~cm$^{-2}$~s$^{-1}$. \citet{Vignati1999} derive a much larger
luminosity by an obscured AGN, with an absorbing screen of $N_{\rm
  H}\sim 10^{24}$~cm$^{-2}$ which also reduces the X-ray fluxes below
$< 1$~erg~cm$^{-2}$~s$^{-1}$. This leaves room for an XDR component
located near the AGN nuclei, but even the most optimistic estimate of
the X-ray luminosity is not enough to explain the combined H$_2$ and
CO luminosity of $L=2.5\times 10^9$~L$_{\odot}=9.6\times
10^{42}$~erg~s$^{-1}$. This would require a strong coupling of the
X-rays to the molecular gas.

\subsection{Shock modeling}\label{modeling}

\begin{figure}
  \centering
  \includegraphics[width=8cm,clip]{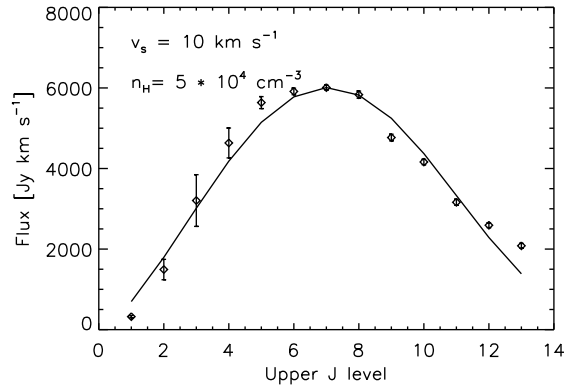}
  \caption{C-shock model (solid line) with density $n_{\rm H}=5\times
    10^4$~cm$^{-3}$ and shock velocity $v_s = 10$~km~s$^{-1}$ overlaid
    on the observed CO line fluxes (diamonds) for NGC 6240.}
  \label{CSHOCK_NGC6240}
\end{figure}

Given our arguments discussed above and the fact that various papers
\citep[cf.,][]{VdWerf1993, Tecza2000} have argued that C type shocks
are exciting the H$_2$ lines in NGC~6240, we analyze the CO ladder
with shock models. Two types of shocks are used: a magnetic continuous
($C$-type) and a non-magnetic jump ($J$-type) shock model. In both
types, ro-vibrational levels of H$_2$ are excited through high
temperature H$_2-$H$_2$, H$-$H$_2$, and He-H$_2$ collisions. As the
shock develops, the temperature of the gas becomes such that the
excited vibrational states become populated. We use the
\citet{Flower2003} shock code to model the H$_2$ $v=1-0\ S(1) / v=2-1\
S(1)$ ratio and the CO ladder. \citet{Kristensen2007} used this model
to calculate a grid, spanning hydrogen number densities between
$n_{\rm H}=10^4$ and $10^7$~cm$^{-3}$, velocities between $v=10$ and
$50$~km~s$^{-1}$, and transverse magnetic field densities $b\times
n^{1/2}_{\rm H}$~$\mu$G, with b=1 and 5. The magnetic field density
relation implies values between 100 and 3000 $\mu$G for the adopted
densities, which is within the range of values observed for galactic
molecular clouds \citep{Crutcher1999}.

Using a rotational diagram and assuming that the CO lines are
optically thin, we find $T_{\rm rot}=66$~K for the CO rotational
transitions between CO $J=5-4$ and $J=7-6$. This rotational
temperature increases to values $T_{\rm rot}\sim 150$~K at the highest
CO transitions. We note that for the highest transitions the
rotational temperature is a lower limit to the kinetic temperature,
since these transitions are slightly sub-thermally excited.  Using a
chi-square fit and including the errors provided in Table
\ref{Fluxes}, the best-fitting C-type shock model has a hydrogen
number density of $n_{\rm H} = 5\times 10^4$~cm$^{-3}$ and velocity of
$v= 10$ km s$^{-1}$ (see Fig. \ref{CSHOCK_NGC6240}). The density is
well constrained, and chi-square values increase by an order of
magnitude when going to densities $n=10^4$ or $10^5$~cm$^{-2}$. The
uncertainty in the shock velocity is a few km~s$^{-1}$. Downstream in
this particular model, the CO gas has been compressed by a factor of
7.2 and the post-shock density of the CO emitting gas is $3.6\times
10^5$ cm$^{-3}$.

The $v=1-0$ $S(1)$ and $v=2-1$ $S(1)$ H$_2$ lines have upper level
energies of $E = 6500$ and 12500~K, respectively, and are therefore
only excited when the gas temperature $T \gtrsim 1000$~K. It turns out
that there is a degeneracy between pre-shock density and velocity for
reproducing the observed $v=1-0\ S(1) / v=2-1\ S(1)$ ratio of $\sim 7$
\citep{Tecza2000}. The lower the pre-shock density, the higher the
velocity that is required: The solutions range between hydrogen number
density $n_{\rm H}=5\times 10^5$~cm$^{-3}$ and a shock velocity
$v_{\rm s} =47$~km\,s$^{-1}$ to a pre-shock density of $n_{\rm H} =
10^{7}$~cm$^{-3}$ combined with a velocity of $v_{\rm s}=16$~km
s$^{-1}$.


\section{Discussion and conclusions}

{\it Shock modeling:} Combining the model results for the H$_2$ and CO
emission, we find that the H$_2$ $v=1-0\ S(1)$ to CO $J=10-9$
intensity ratio is of order $\sim 100$. In contrast, the observed
H$_2$ / CO line luminosity ratio is approximately $\sim 0.5$. The low
density, low velocity shock model fitting the CO lines, has a low
temperature and does not produce H$_2$ emission. From this we conclude
that only a very small fraction of the gas mass is currently exposed
to very powerful shocks (with either very high densities, $n_{\rm
  H}=10^7$~cm$^{-3}$ and moderate shock velocities $v_{\rm
  s}=16$~km\,s$^{-1}$ or moderate densities, $n_{\rm H}=4\times
10^4$~cm$^{-3}$, combined with a high shock velocity $v_{\rm s} \sim
~50$~km~s$^{-1}$). Most of the shocked gas is settling and
equilibrating with the ambient ISM, which is in agreement with the
fast dissipation timescale derived below.

{\it Dissipation timescales:} If we assume that all the gas is
colliding at a shock velocity $v_s = 50$~km~s$^{-1}$ (which is the
highest velocity allowed by the models reproducing the H$_2$ lines),
the total amount of energy available (for an adopted gas mass of
$M=1.6\times 10^{10}$~M$_\odot$, in the middle of the LVG masses
derived by various authors) is $E=0.5\ M\ v_s^2 = 4.0\times
10^{56}$~erg, which is very similar to the value derived by
\cite{Tacconi1999}. Assuming no additional energy input and that CO
traces the bulk of the gas mass, this would imply that the CO shock
energy is dissipated away within 6.6 million years, well within the
orbital timescale of the two nuclei of $30$ million years
\citep{Tacconi1999}.

{\it Excitation by UV/X-rays vs. Shocks:} The observed H$_2$ $v=1-0\
S(1)$ / $v=2-1\ S(1)$ ratio falls within the range of ratios that are
produced by X-ray dominated regions \citep[see,
e.g.,][Fig. 6]{Maloney1996}. Also, the CO ladder of NGC 6240 resembles
the one observed for Mrk 231 \citep{VdWerf2010}. However, as mentioned
before, NGC 6240 is lacking the bright OH$^+$ and H$_2$O$^+$ lines,
which are associated with gas clouds that are exposed to extremely
high cosmic ray or X-ray ionization rates \citep{Meijerink2011}, and
the available X-ray photons are not sufficient to dominate the
chemistry and thermal balance of the bulk of the gas (see
Sect. \ref{modeling}). The H$_2$O lines are less luminous than in Mrk
231. This implies either lower water abundances or a less efficient
mode of excitation. A full analysis is beyond the scope of this
paper. These results and the similarity to the high-$J$ CO line
distribution in Mrk 231 suggest that shocks possibly due to the
massive molecular outflow \citep{Fischer2010,Feruglio2010,Sturm2011}
may also contribute to the CO line emission in Mrk 231. Indeed, based
on millimeter interferometric observations of Mrk 231,
\citet{Cicone2012} note that in the inner region ($R < 0.3$~kpc), the
CO($2-1$)/CO($1-0$) ratio is slightly higher, indicative of a shock
contribution.
 
{\it Observing strategies for ALMA and other sub-millimeter
  facilities:} Although the observed CO ladders for Mrk 231 and NGC
6240 are practically indistinguishable, we argue that shock excitation
and not X-rays are responsible for the excitation of the CO ladder,
based on our knowledge of the geometry of the NGC 6240. Such an
analysis will not be possible in the study of the ISM in galaxies at
high redshift. At those distances, we are unable to resolve the offset
between the two radio nuclei and the location of the CO emitting gas,
and X-ray fluxes cannot be determined. Therefore, we have to rely on
dust continuum and line emission at far-infrared and sub-millimeter
wavelengths. {\it Here we have argued that a high line-to-continuum
  ratio is a key diagnostic for the presence of shocks}. Additionally,
more detailed modeling of emission by molecules and ions, such as
H$_2$O, H$_2$O$^+$ and OH$^+$, will help in making this distinction,
and are thus also highly recommended to include in future ALMA
programs.

\begin{acknowledgements}
  We acknowledge the constructive comments by the referee Christine
  Wilson. The following institutes have provided hardware and software
  elements to the SPIRE project: University of Lethbridge, Canada;
  NAOC, Beijing, China; CEA Saclay, CEA Grenoble and LAM in France;
  IFSI, Rome, and University of Padua, Italy; IAC, Tenerife, Spain;
  Stockholm Observatory, Sweden; Cardiff University, Imperial College
  London, UCL-MSSL, STFCRAL, UK ATC Edinburgh, and the University of
  Sussex in the UK. Funding for SPIRE has been provided by the
  national agencies of the participating countries and by internal
  institute funding: CSA in Canada; NAOC in China; CNES, CNRS, and CEA
  in France; ASI in Italy; MEC in Spain; Stockholm Observatory in
  Sweden; STFC in the UK; and NASA in the USA. Additional funding
  support for some instrument activities has been provided by ESA. US
  authors ackowledge support from the NHSC. Basic research in IR
  astronomy at NRL is funded by the US ONR.
\end{acknowledgements}


\end{document}